# Erasure-Coded Byzantine Storage with Separate Metadata

Elli Androulaki*    Christian Cachin*    Dan Dobre†    Marko Vukolić‡

3 May 2018

**Abstract**

Although many distributed storage protocols have been introduced, a solution that combines the strongest properties in terms of availability, consistency, fault-tolerance, storage complexity and the supported level of concurrency, has been elusive for a long time. Combining these properties is difficult, especially if the resulting solution is required to be efficient and incur low cost.

We present AWE, the first *erasure-coded* distributed implementation of a multi-writer multi-reader read/write storage object that is, at the same time: (1) asynchronous, (2) wait-free, (3) atomic, (4) amnesic, (i.e., with data nodes storing a bounded number of values) and (5) Byzantine fault-tolerant (BFT) using the optimal number of nodes. Furthermore, AWE is efficient since it does not use public-key cryptography and requires data nodes that support only reads and writes, further reducing the cost of deployment and ownership of a distributed storage solution. Notably, AWE stores metadata separately from $k$-out-of-$n$ erasure-coded fragments. This enables AWE to be the first BFT protocol that uses as few as $2t + k$ data nodes to tolerate $t$ Byzantine nodes, for any $k \geq 1$.

## 1  Introduction

**Background.** *Erasure coding* is a key technology that saves space and retains robustness against faults in distributed storage systems. In short, an erasure code splits a large data object into $n$ fragments such that from any $k$ of them the input value can be reconstructed. The utility of erasure coding is demonstrated by large-scale erasure-coding storage systems that have been deployed today [22, 27]. These distributed storage systems offer large capacity, high throughput, and resilience to faults.

Whereas the storage systems in production use today only tolerate component crashes or outages, storage systems in the *Byzantine failure model* survive also more severe faults, ranging from arbitrary state corruption to malicious attacks on components. In this paper, we consider a model where *clients* directly access a storage service provided by distributed servers, called *nodes* — a fraction of the nodes may be Byzantine, whereas clients may fail as well, but only by crashing.

Although Byzantine-fault tolerant (BFT) erasure-coded distributed storage systems have received some attention in the literature [4, 9, 14, 16, 19], our understanding of their properties lies behind that of replicated storage. In fact, most existing BFT erasure-coded storage approaches have drawbacks that prevented their wide-spread use. For example, they relied on the nodes storing an unbounded number of values [16], required the nodes to communicate with each other [9], used public-key cryptography [9, 19], or might have blocked clients due to concurrent operations of other clients [19].

We consider an abstract *wait-free* storage register with *atomic* semantics [21], accessed concurrently by multiple readers and writers (MRMW). Wait-free termination means that any client operation terminates irrespective of the behavior of the Byzantine nodes and of other clients. This is not easy to achieve with Byzantine nodes [1] even in systems that replicate the data. Therefore, previous works have often used a weaker notion of liveness called *finite-write (FW) termination*, which ensures that read operations progress only in executions with a finite number of writes.

*IBM Research - Zurich, Rüschlikon, Switzerland, {lli,cca}@zurich.ibm.com.
†NEC Labs Europe, Germany, dan.dobre@neclab.eu.
‡Eurécom, Sophia Antipolis, France, vukolic@eurecom.fr.

**Contribution.** This paper introduces AWE, the *first* asynchronous, wait-free distributed BFT erasure-coded storage protocol with optimal resilience. As in previous work, we assume there are $n$ nodes and up to $t$ of them may exhibit non-responsive (NR-)arbitrary faults, that is, Byzantine corruptions. The best resilience that has been achieved so far is $n > 3t$, which is optimal for Byzantine storage [24]. However, our protocol features a separation of metadata and erasure coded fragments; with this approach our protocol may reduce the number of *data nodes*, i.e., those that store a fragment, to lower values than $n$ for $k \leq t$. In particular, our protocol takes only $2t + k$ data nodes; this idea saves resources, as in the separation of agreement and execution for BFT services [28]. For implementing the metadata service, $n > 3t$ nodes are still needed.

Our protocol employs simple, passive data nodes; they cannot execute code and they only support read and write operations, such as the key-value stores (KVS) provided by popular cloud storage services. The metadata service itself is an atomic snapshot object, which has only weak semantics and may be implemented in a replicated asynchronous system from simple read/write registers [3]. The protocol is also *amnesic* [11], i.e., the nodes store a bounded number of values and may erase obsolete data. The protocol uses only simple cryptographic hash functions but no (expensive) public-key operations.

In summary, protocol AWE, introduced in Section 3, is the first erasure-coded distributed implementation of a MRMW storage object that is, at the same time: (1) asynchronous, (2) wait-free, (3) atomic, (4) amnesic, (5) tolerates the optimal number of Byzantine nodes, and (6) does not use public-key cryptography. Furthermore, AWE can be implemented from non-programmable nodes (KVS) that only support reads and writes (in the vein of Disk Paxos [1]). In practice, the KVS interface is offered by commodity cloud storage services, which could be used as AWE data nodes to reduce the cost of AWE deployment and ownership. While some of these desirable properties have been achieved in different combinations so far, they have never been achieved together with erasure-coded storage, as explained later. Combining these properties has been a longstanding open problem [16].

**Related work.** We provide a brief overview of the most relevant literature on the subject. Table 1 summarizes this section.

Earlier designs for erasure-coded distributed storage have suffered from potential aborts due to contention [15] or from the need to maintain an unbounded number of fragments at data nodes [16]. In the crash-failure model, ORCAS [14] and CASGC [10] achieve optimal resilience $n > 2t$ and low communication overhead, combined with wait-free (ORCAS) and FW-termination (CASGC), respectively.

In the model with Byzantine nodes, Cachin and Tessaro (CT) [9] introduced the first wait-free protocol with atomic semantics and optimal resilience $n > 3t$. CT uses a verifiable information dispersal protocol but needs node-to-node communication, which lies outside our model. Hendricks et al. (HGR) [19] present an optimally resilient protocol that comes closest to our protocol among the existing solutions. It offers many desirable features, that is, it has as low communication cost, works asynchronously, achieves optimal resilience, atomicity, and is amnesic. Compared to our work, it uses public-key cryptography, achieves only FW-termination instead of wait-freedom, and requires *processing* by the nodes, i.e., the ability to execute complex operations beyond simple reads and writes.

To be fair, much of the (cryptographic) overhead inherent in the CT and HGR protocols defends against poisonous writes from Byzantine clients, i.e., malicious client behavior that leaves the nodes in an inconsistent state. We do not consider Byzantine clients in this work, since permitting arbitrary client behavior is problematic. Such a client might write garbage to the storage system at any time and wipe out the stored value. Furthermore, the standard formal correctness notions such as linearizability fail when clients misbehave (apart from crashing). Hendricks [18] discusses correctness notions in the presence of Byzantine clients. However, even without the steps that protect against poisonous writes, HGR still requires processing by the nodes and is not wait-free.

| Protocol | BFT | Liveness | Data nodes | Type | Amnesic | Cryptogr. |
|---|---|---|---|---|---|---|
| ORCAS [14] | — | Wait-free | $2t+1$ | Proc. | — | N/A |
| CASGC [10] | — | FW-term. | $2t+1$ | Proc. | ✓* | N/A |
| CT [9] | ✓* | Wait-free * | $3t+1$ | Msg. | — | Public-key |
| HGR [19] | ✓* | FW-term. | $2t+k$, for $k>t$ | Proc. | ✓* | Public-key |
| M-PoWerStore [12] | ✓* | Wait-free * | $3t+1$ | Proc. | — | Hash func. * |
| DepSky [4] | ✓* | Obstr.-free | $3t+1$ | R/W * | — | Public-key |
| AWE (Sec. 3) | ✓* | Wait-free * | $2t+k$, for $k\geq 1$ * | R/W * | ✓* | Hash func. * |

Table 1: Comparison of erasure-coded distributed storage solutions. An asterisk (*) denotes optimal properties. The column labeled *Type* states the computation requirements on nodes: *Proc.* denotes processing; *Msg.* means sending messages to other nodes, in addition to processing; *R/W* means a register object supporting only read and write.

The M-PoWerStore protocol [12] employs a cryptographic "proof of writing" for wait-free atomic erasure-coded distributed storage. It is the first wait-free BFT solution without node-to-node communication. Similar to other protocols, M-PoWerStore uses nodes with processing capabilities and is not amnesic.

Several systems have recently addressed how to store erasure-coded data on multiple redundant cloud services but only few of them focus on wait-free concurrent access. HAIL [5], for instance, uses Byzantine-tolerant erasure coding and provides data integrity through proofs of retrievability; however, it does not address concurrent operations by different clients. DepSky [4] achieves regular semantics and uses lock-based concurrency control; therefore, one client may block operations of other clients.

A key aspect of AWE lies in the differentiation of (small) metadata from (large) bulk data: this enables a modular protocol design and an architectural separation for implementations. The FARSITE system [2] first introduced such a separation for replicated storage; their data nodes and their metadata abstractions require processing, however, in contrast to AWE. Non-explicit ways of separating metadata from data can already be found in several previous erasure coding-based protocols. For instance, the cross checksum, a vector with the hashes of all $n$ fragments, has been replicated on the data nodes to ensure consistency [9, 16].

Finally, a recent protocol called MDStore [6] has shown that separating metadata from bulk data permits to reduce the number of data nodes in asynchronous wait-free BFT distributed storage implementations to only $2t+1$. When protocol AWE is reduced to use replication with the trivial erasure code ($k=1$), it uses as few nodes as MDStore to achieve the same wait-free atomic semantics; unlike AWE, however, MDStore is not amnesic and uses processing nodes.

**Structure.** The paper continues with the model in Section 2 and presents Protocol AWE in Section 3. The communication and storage complexities of AWE are compared to those of existing protocols in Section 4. Section 5 contains a formal proof for the properties of AWE.

## 2 Definitions

**System model.** We consider an asynchronous distributed system of components (or processes) that communicate with each other. The components contain a set $\mathcal{C}$ of $m$ *clients*, a set $\mathcal{D}$ of $n$ *data nodes* $d_1, \ldots, d_n$, and further process abstractions. The components interact asynchronously via exchanging events. A protocol specifies a collection of programs with instructions for all components.

A component may fail by crashing or by exhibiting *Byzantine* faults; the latter means they may deviate arbitrarily from their specification. We assume that clients can only crash; on the other hand, up to $t$ data nodes can be Byzantine and behave adversarially (NR-arbitrary faults). A component that does not fail is called *correct*.

**Notation.** Protocols are presented in a modular way using an event-based notation [7]. A component is specified through its *interface*, containing the events that it exposes to other components that may call it, and its *properties*, which define its behavior. A component may react to a received event by doing computation and triggering further events. Every component is named by an identifier. Events are qualified by the component identifier to which the event belongs and may take parameters. An event *Sample* of a component $m$ with a parameter $x$ is denoted by $\langle\, m\text{-}Sample \mid x \,\rangle$.

Components interact asynchronously with others through exchanging events. We assume that all events communicated from one component to another are delivered in FIFO-order. There are two kinds of events in a component's interface: *input events* that it receives from other components, typically to invoke its services, and *output events*, through which the component delivers information or signals a condition to another component. The behavior of a component is typically stated through a number of properties or through a sequential implementation.

**Objects and histories.** An *object* is a special type of component for which every input event (called an *invocation* in this context) triggers exactly one output event (called a *response*). Every such pair of invocation and response define an *operation* of the object. An operation *completes* when its response occurs.

A *history* $\sigma$ of an execution of an object $O$ consists of the sequence of invocations and responses of $O$ occurring in $\sigma$. An operation is called *complete* in a history if it has a matching response. An operation $o$ *precedes* another operation $o'$ in a sequence of events $\sigma$, denoted $o <_\sigma o'$, whenever $o$ completes before $o'$ is invoked in $\sigma$. If $o$ precedes $o'$ then $o'$ *follows* $o$. A sequence of events $\pi$ *preserves the real-time order* of a history $\sigma$ if for every two operations $o$ and $o'$ in $\pi$, if $o <_\sigma o'$ then $o <_\pi o'$. Two operations are *concurrent* if neither one of them precedes the other. A sequence of events is *sequential* if it does not contain concurrent operations. We often simplify the terminology by exploiting that every *sequential* sequence of events corresponds naturally to a sequence of operations.

An execution is *well-formed* if the events at every object are alternating invocations and matching responses, starting with an invocation. An execution is *fair*, informally, if it does not halt prematurely when there are still steps to be taken or triggered events to be consumed (see the standard literature for a formal definition [23]).

**Registers.** A *read/write register* $r$ is an object that stores a value from a domain $\mathcal{V}$ and supports exactly two operations, for writing and reading the value. More precisely:

- A *Write* operation to $r$ is triggered by an invocation $\langle\, r\text{-}Write \mid v \,\rangle$ that takes a value $v \in \mathcal{V}$ as parameter and terminates by generating a response $\langle\, r\text{-}WriteAck \,\rangle$ with no parameter.

- A *Read* operation from $r$ is triggered by an invocation $\langle\, r\text{-}Read \,\rangle$ with no parameter; the register signals that the read operation completes by triggering a response $\langle\, r\text{-}ReadResp \mid v \,\rangle$, which contains a parameter $v \in \mathcal{V}$.

The behavior of a register is given through its sequential specification, which requires that every *r-Read* operation returns the value written by the last preceding *r-Write* operation in the execution, or the special symbol $\perp \notin \mathcal{V}$ if no such operation exists. For simplicity, we will assume that every distinct value is written only once.

In this work, any client may invoke the operations of the emulated register object; such registers are also called *multi-reader multi-writer (MRMW) registers*. Furthermore, we assume that all clients invoke a well-formed sequence of operations.

**Consistency and availability.** Recall that clients interact with an object $O$ through its operations, defined in terms of an invocation and a response event of $O$. We say that a client $c$ *executes* an operation between the corresponding invocation and response events. When accessed concurrently by multiple processes, executions of objects considered in this work are *linearizable*, that is, the object appears to execute all operations *atomically*.

**Definition 1 (View).** A sequence of events $\pi$ is called a *view* of a history $\sigma$ at a client $c$ w.r.t. an object $O$ whenever:

1. $\pi$ is a sequential permutation of some subsequence of complete operations in $\sigma$;
2. all complete operations executed by $c$ appear in $\pi$; and
3. $\pi$ satisfies the sequential specification of $O$.

**Definition 2 (Linearizability [21]).** A history $\sigma$ is linearizable w.r.t. an object $O$ if there exists a sequence of events $\pi$ such that:

1. $\pi$ is a view of $\sigma$ at all clients w.r.t. $O$; and
2. $\pi$ preserves the real-time order of $\sigma$.

The goal of this work is to describe a protocol that emulates a linearizable register abstraction among the clients; such a register is also called *atomic*. Some of the clients may crash and some nodes may be Byzantine, but every client operation should terminate in all cases, irrespective of how other clients and nodes behave.

**Definition 3 (Wait-freedom [20]).** A protocol is called *wait-free* if every operation invoked by a correct client eventually completes.

**Cryptography.** We make use of cryptographic hash functions. One can imagine that the cryptographic schemes are implemented by a distributed oracle accessible to all components [7]. A hash function $H$ maps a bit string $x$ of arbitrary length to a short, unique representation of fixed length. We use a *collision-free* hash function; this property means that no process, not even a Byzantine component, can find two distinct values $x$ and $x'$ such that $H(x) = H(x')$.

## 3 Protocol AWE

This section introduces the *asynchronous wait-free erasure-coded Byzantine distributed storage protocol (AWE)*.

### 3.1 Abstractions

**Erasure code.** An $(n,k)$-*erasure code (EC)* with domain $\mathcal{V}$ is given by an encoding algorithm, denoted *Encode*, and a reconstruction algorithm, called *Reconstruct*. Given a (large) value $v \in \mathcal{V}$, algorithm $Encode_{k,n}(v)$ produces a vector $[f_1, \ldots, f_n]$ of $n$ *fragments*, which are from a domain $\mathcal{F}$. A fragment is typically much smaller than the input, and any $k$ fragments contain all information of $v$, that is, $|\mathcal{V}| \approx k|\mathcal{F}|$.

For an $n$-vector $F \in (\mathcal{F} \cup \{\bot\})^n$, whose entries are either fragments or the symbol $\bot$, algorithm $Reconstruct_{k,n}(F)$ outputs a value $v \in \mathcal{V}$ or $\bot$. An output value of $\bot$ means that the reconstruction failed. The *completeness* property of an erasure code requires that an encoded value can be reconstructed from any $k$ fragments. In other words, for every $v \in \mathcal{V}$, when one computes $F \leftarrow Encode_{k,n}(v)$ and then erases up to $n - k$ entries in $F$ by setting them to $\bot$, algorithm $Reconstruct_{k,n}(F)$ outputs $v$. More details are available in the literature [25, 26].

**Metadata service.** The metadata service is implemented by a standard *atomic snapshot object* [3], called *dir*, that serves as a *directory*. A snapshot object extends the simple storage function of a register to a service that maintains one value for each client and allows for better coordination. Like an array of multi-reader single-writer (MRSW) registers, it allows every client to *update* its value individually; for reading it supports a *scan* operation that returns the vector of the stored values, one for every client. More precisely, the operations of *dir* are:

- An *Update* operation to *dir* is triggered by an invocation $\langle$ *dir-Update* $\mid c, v$ $\rangle$ by client $c$ that takes a value $v \in \mathcal{V}$ as parameter and terminates by generating a response $\langle$ *r-UpdateAck* $\rangle$ with no parameter.

- A *Scan* operation on *dir* is triggered by an invocation $\langle$ *dir-Scan* $\rangle$ with no parameter; the snapshot object returns a vector $V$ of $m = |\mathcal{C}|$ values to $c$ as the parameter in the response $\langle$ *r-ScanResp* $\mid V$ $\rangle$, with $V[c] \in \mathcal{V}$ for $c \in \mathcal{C}$.

The sequential specification of the snapshot object follows directly from the specification of an array of $m$ MRSW registers (hence, the snapshot initially stores the special symbol $\bot \notin \mathcal{V}$ in every entry). When accessed concurrently from multiple clients, its operations appear to take place atomically, i.e., they are linearizable. Snapshot objects are weak — they can be implemented from read/write registers [3], which, in turn, can be implemented from a set of a distributed processes subject to Byzantine faults. Wait-free amnesic implementations of registers with the optimal number of $n > 3t$ processes are possible using existing constructions [13, 17].

### 3.2 Protocol overview

The high-level architecture of AWE uses the metadata directory *dir* to maintain pointers to the fragments stored at the data nodes. As in standard implementations of multi-writer distributed storage [7], every value is associated to a timestamp, which consists of a sequence number *sn* and the identifier $c$ of the writing client, i.e., $ts = (sn, c) \in \textit{Timestamps} = \mathbb{N}_0 \times (\mathcal{C} \cup \{\bot\})$; timestamps are initialized to $T_0 = (0, \bot)$. The metadata contains the timestamp of the most recently written value for every client, and readers determine the value to read by retrieving all timestamps, determining their maximum, and accessing the fragments associated to the highest timestamp. Comparisons among timestamps use the standard ordering, where $ts_1 > ts_2$ for $ts_1 = (sn_1, c_1)$ and $ts_2 = (sn_2, c_2)$ if and only if $sn_1 > sn_2 \vee (sn_1 = sn_2 \wedge c_1 > c_2)$.

The directory stores an entry for every writer; it contains the timestamp of its most recently written value, the identities of those nodes that have acknowledged to store a fragment of it, a vector with the hashes of the fragments for ensuring data integrity, and additional metadata to support concurrent reads and writes. The linearizable semantics of protocol AWE are obtained from the atomicity of the metadata directory.

At a high level, the writer first invokes *dir-Scan* on the metadata to read the highest stored timestamp, increments it, and uses this as the timestamp of the value to be written. Then it encodes the value to $n$ fragments and sends one fragment to each data node. The data nodes store it and acknowledge the write. After the writer has received acknowledgments from $t + k$ data nodes, it writes their identities (together with the timestamp and the hashes of the fragments) to the metadata through *dir-Update*. The reader proceeds accordingly: it first invokes *dir-Scan* to obtain the entries of all writers; it determines the highest timestamp among them and extracts the fragment hashes and the identities of the data nodes; finally, it contacts the data nodes and reconstructs the value after obtaining $k$ fragments that match the hashes in the metadata.

Although this simplified algorithm achieves atomic semantics, it does not address timely garbage-collection of obsolete fragments, the main problem to be solved for amnesic erasure-code distributed storage. It is easy to see that overwriting the fragments during the next write operation may cause a reader to stall.

Protocol AWE uses two mechanisms to address this: first, the writer *retains* those values that may be accessed concurrently and exempts them from garbage collection so that their fragments remain intact for concurrent readers, which gives the reader enough time to retrieve its fragments. Secondly, some of the retained values may also be *frozen* in response to concurrent reads; this forces a concurrent read to retrieve a value that is guaranteed to exist at the data nodes rather than simply the newest value, thereby effectively limiting the amount of stored values. A similar freezing method has been used for wait-free atomic storage with replicated data [13, 17], but it must be changed for erasure-coded storage with separated metadata. The retention technique together with the separation of metadata appears novel.

For the two mechanisms, every reader maintains a *reader index*, both in its local variable *readindex* and in its metadata. The reader index serves for coordination between the reader and the writers. The reader increments its index whenever it starts a new *r-Read* and immediately writes it to *dir*, thereby announcing its intent to read. Writers access the reader indices after updating the metadata for a write and before (potentially) erasing obsolete fragments. Every writer $w$ maintains a table *frozenindex* with its most recent recollection of all reader indices. When the newly obtained index of a reader $c$ has changed, then $w$ detects that $c$ has started a new operation at some time after the last write of $w$.

When $w$ detects a new operation of $c$, it does not know whether $c$ has retrieved the timestamp from *dir* before or after the *dir-Update* of the current write. The reader may access either value; the writer therefore *retains* both the current and the preceding value for $c$ by storing a pointer to them in *frozenptrlist* and in *reservedptrlist*. Clearly, both values have to be excluded from garbage collection by $w$ in order to guarantee that the reader completes.

However, the operation of the reader $c$ may access *dir* after the *dir-Update* of one or more subsequent write operation by $w$, which means that the nodes would have to retain every value subsequently written by $w$ as well. To prevent this from happening and to limit the number of stored values, $w$ *freezes* the currently written timestamp (as well as the value) and forces $c$ to read this timestamp when it accesses *dir* within the same operation. In particular, the writer stores the current timestamp in *frozenptrlist* at index $c$ and updates the reader index of $c$ in *frozenindex*; then, the writer pushes both tables, *frozenindex* and *frozenptrlist*, to the metadata service during its next *r-Write*. The values designated by *frozenptrlist* (they are called *frozen*) and *reservedptrlist* (they are called *reserved*) are retained and excluded from garbage collection until $w$ detects the next read of $c$, i.e., the reader index of $c$ increases. Thus, the current read may span many concurrent writes of $w$ and the fragments remain available until $c$ finishes reading.

On the other hand, a reader must consider frozen values. When a slow read operation spans multiple concurrent writes, the reader $c$ learns that it should retrieve the frozen value through its entry in the *frozenindex* table of the writer. More precisely, when $c$ retrieves the metadata from *dir* and finds that writer $w$'s *frozenindex*[$c$] entry equals its *readindex* variable, then $w$ has frozen the value designated by *frozenptrlist*[$c$] for $c$.

The protocol is amnesic because each writer retains at most two values per reader, a frozen value and a reserved value. Every data node therefore stores at most two fragments for every reader-writer pair plus the fragment from the currently written value. The combination of freezing and retentions ensures that readers never wait.

## 3.3 Details

**Data structures.** We use abstract data structures for compactness. In particular, given a timestamp $ts = (sn, c)$, its two fields can be accessed as $ts.sn$ and $ts.c$. A data type *Pointers* denotes a set of tuples of the form $(ts, set, hash)$ with $ts \in \textit{Timestamps}$, $set \subseteq [1, n]$, and $hash[i] \in \Sigma^*$ for $i \in [1, n]$. Their initialization value is $\textit{Nullptr} = ((0, \bot), \emptyset, [\bot, \ldots, \bot])$.

A *Pointers* structure contains the relevant information about one stored value. For example, the writer locally maintains $writeptr \in \textit{Pointers}$ designating to the most recently written value. More specifically, *writeptr.ts* contains the timestamp of the written value, *writeptr.set* contains the identities of the nodes that have confirmed to have stored the written value, and *writeptr.hash* contains the cross checksum, the list of hash values of the data fragments, of the written value.

The metadata directory *dir* contains a vector $M$ with a tuple for every client $p \in \mathcal{C}$ of the form

$$M[p] = (\textit{writeptr}, \textit{frozenptrlist}, \textit{frozenindex}, \textit{readindex}),$$

where the field $writeptr \in \textit{Pointers}$ represents the *written value*, the field *frozenptrlist* is an array indexed by $c \in \mathcal{C}$ such that $frozenptrlist[c] \in \textit{Pointers}$ denotes a value *frozen by p for reader c*, and the integer *readindex* denotes the reader-index of $p$.

For preventing that concurrently accessed fragments are garbage-collected, the writer maintains two tables, *frozenptrlist*, and *reservedptrlist*, each containing one *Pointers* entry for every reader in $\mathcal{C}$. The second one, *reservedptrlist*, is stored only locally, together with the *frozenindex* table, which denotes the writer's most recently obtained copy of the reader indices. For the operations of the reader, only the local *readindex* counter is needed.

Every client maintains the following variables between operations: *writeptr*, *frozenptrlist*, *frozenindex*, and *reservedptrlist* implement freezing, reservations, and retentions for writers as mentioned, and *readindex* counts the reader operations.

When clients access *dir*, they may not be interested to retrieve all fields or to update all fields. For clarity we replace the fields to be ignored by $*$ in those *dir-Scan* and *dir-Update* operations.

**Operations.** At the start of a write operation, the writer $w$ saves the current value of *writeptr* in *prevptr*, to be used later during its operation, if $w$ should reserve and retain that value. Then $w$ determines the timestamp of the current operation, which is stored in *writeptr.ts*. After computing the fragments of $v$, sending them to the data nodes, and obtaining $t + k$ acknowledgements, the writer updates its metadata entry. It writes *writeptr*, pointing to $v$, together with *frozenptrlist* and *frozenindex*, as they resulted after the previous write to *dir*. Then $w$ invokes *dir-Scan* and acquires the current metadata $M$, which it uses to determine values to freeze and to retain. It compares the acquired reader indices with the ones obtained during its last write (as stored in *frozenindex*). When $w$ detects a read operation by $c$ because $M[c].readindex > frozenindex[c]$, it freezes the current value (by setting $frozenptrlist[p]$ to *writeptr*) and reserves the previously written value (by setting $reservedptrlist[p]$ to *prevptr*). Finally, the writer deletes all fragments at the data nodes except for those of the currently written and the retained values.

To determine the timestamps for retrieving fragments, the reader uses the following two functions:

**function** $\textit{readfrom}(M, c, p, \textit{index})$ **is**
    **if** $\textit{index} > M[p].\textit{frozenindex}[c]$ **then**
        **return** $M[p].\textit{writeptr}$
    **else**   // $\textit{index} = M[p].\textit{frozenindex}[c]$
        **return** $M[p].\textit{frozenptrlist}[c]$

**function** $\textit{highestread}(M, c, \textit{index})$ **is**
    $max \leftarrow \textit{Nullptr}$
    **forall** $p \in \mathcal{C}$ **do**
        $ptr \leftarrow \textit{readfrom}(M, c, p, \textit{index})$
        **if** $ptr.ts > max.ts$ **then**
            $max \leftarrow ptr$
    **return** $max$

Upon retrieving the array $M$ from *dir*, the reader sets *readptr* $\leftarrow$ *highestread*$(M, c, readindex)$, which implements the logic of accessing frozen timestamps. The two functions above ensure that

$$\begin{aligned}
\text{readfrom}(M, c, p, \text{index}) = & \\
\big(\text{ptr} \in \text{Pointers} \;: & \\
& (\text{ptr} = M[p].\text{writeptr} \wedge \text{index} > M[p].\text{frozenindex}[c]) \\
& \vee (\text{ptr} = M[p].\text{frozenptrlist}[c] \wedge \text{index} = M[p].\text{frozenindex}[c])\big)
\end{aligned}$$

$$\begin{aligned}
\text{highestread}(M, c, \text{index}) = & \\
& \arg\max_{\text{ptr} \in \text{Readset}}\{\text{ptr.ts}\}, \text{ where } \text{Readset} = \{\text{readfrom}(M, c, p, \text{index}) \mid p \in \mathcal{C}\}
\end{aligned}$$

The details of protocol AWE appear in Algorithms 1–3.

**Remarks.** Note that AWE does not need a majority of correct data nodes and neither refers to quorum systems for correctness; these aspects are all encapsulated in the directory service. For liveness, though, the protocol needs to obtain responses from $t + k$ data nodes during write operations, which is only possible if $n \geq 2t + k$.

In the current formulation of AWE, every writer retains exactly two values for each reader, regardless of whether the reader has completed its operation. In fact, a value continues to be retained for a reader $c$ until $c$ invokes a subsequent *r-Read* (and concurrently or later, the writer invokes another *r-Write*). In order to avoid retaining unnecessary values, one could introduce an additional field in the metadata for each reader, through which the reader can signal when it completes a read operation. The writer would periodically check this and remove the values no longer needed.

The data nodes can be implemented from a key-value store (KVS) abstraction that has become a prominent interface for cloud-storage systems. A KVS can be implemented from read/write registers, as shown by Cachin et al. [8], though their implementation does not preserve the space complexity.

## 4 Complexity comparison

This section compares the communication and storage complexities of AWE to existing erasure-coded distributed storage solutions, in a setting with $n$ data nodes and $m$ clients. We denote the size of each stored value $v \in \mathcal{V}$ by $\ell = \lceil \log_2 |\mathcal{V}| \rceil$. In line with the intended deployment scenarios, we assume that $\ell$ is much larger (by several orders of magnitude) than $n^2$ and $m^2$, i.e., $\ell \gg n^2$ and $\ell \gg m^2$.

We examine the worst-case communication and storage costs incurred by a client in the protocol and distinguish metadata operations (on *dir*) from operations on the data nodes with data (i.e., erasure-coded fragments of data values).

For protocol AWE, the metadata of one value written to *dir* consists of a pointer, containing the cross checksum with $n$ hash values, the $t + k$ identities of the data nodes that store a data fragment, and a timestamp. Moreover, the metadata entry of one writer contains also the list of $m$ pointers to frozen values, the $m$ indices relating to the frozen values, and the writer's reader index. Assuming a collision-resistant hash function with output size $\lambda$ bits and timestamps no larger than $\lambda$ bits, the total size of the metadata is $O(m^2 n \lambda)$. (Note that a $2^\lambda$-bit counter suffices for all protocol executions where the hash function is secure, as collisions in hash functions can be found with about $2^{\lambda/2}$ operations.) In the remainder of this section, the size of the metadata is considered to be negligible and is ignored, though it would incur in practice.

According to the above assumption, the complexity of AWE is dominated by the data itself. When writing a value $v \in \mathcal{V}$, the writer sends a fragment of size $\ell/k$ and a timestamp of size $\lambda$ to each of the $n$ data nodes. Assuming further that $\ell \gg \lambda$, the total storage space occupied by $v$ at the data nodes amounts to $n\ell/k$ bits. Similarly, a read operation incurs a communication cost of $(t + k)k/\ell$ bits.

**Algorithm 1.** Protocol AWE, atomic register instance $r$ for client $c$ (part 1).

**Uses**
    Atomic snapshot object, **instance** $dir$
    Data nodes, **instances** $d_1, \ldots, d_n$

**State**
    // State maintained across write and read operations
    *writeptr* $\in$ *Pointers*, initially *Nullptr*         // Metadata of the currently written value
    *frozenptrlist*$[p]$ $\in$ *Pointers*, initially *Nullptr*, for $p \in \mathcal{C}$   // Value frozen and retained for reader $p$
    *reservedptrlist*$[p]$ $\in$ *Pointers*, initially *Nullptr*, for $p \in \mathcal{C}$   // Value reserved and retained for reader $p$
    *frozenindex*$[p]$ $\in \mathbb{N}_0$, initially 0, for $p \in \mathcal{C}$     // Last known reader index of $p$
    *readindex* $\in \mathbb{N}_0$, initially 0         // Reader index of $c$
    // Temporary state during operations
    *prevptr* $\in$ *Pointers*, initially *Nullptr*         // Metadata of the value written by $c$ prior to current write
    *readptr* $\in$ *Pointers*, initially *Nullptr*         // Metadata of the value to be read by $c$
    *readlist*$[i]$ $\in \Sigma^*$, initially $\bot$, for $i \in [1, n]$   // List of nodes that have responded during read

**upon** $\langle$ *r-Write* $\mid v$ $\rangle$ **do**
    *prevptr* $\leftarrow$ *writeptr*
    **invoke** $\langle$ *dir-Scan* $\rangle$; **wait for** $\langle$ *dir-ScanResp* $\mid M$ $\rangle$
    $(wsn, *) \leftarrow \max\{M[p].writeptr.ts \mid p \in \mathcal{C}\}$   // Highest timestamp field $ts$ in a *writeptr* in $M$
    *writeptr.ts* $\leftarrow (wsn + 1, c)$         // Construct metadata of the currently written value
    *writeptr.set* $\leftarrow \emptyset$
    $[v_1, \ldots, v_n] \leftarrow Encode_{k,n}(v)$
    **forall** $i \in [1, n]$ **do**
        *writeptr.hash*$[i] \leftarrow H(v_i)$
        **invoke** $\langle$ $d_i$-*Write* $\mid$ *writeptr.ts*, $v_i$ $\rangle$

**upon** $\langle$ $d_i$-*WriteAck* $\mid$ *ats* $\rangle$ **such that** *ats* = *writeptr.ts* $\wedge$ $|writeptr.set| < t + k$ **do**
    *writeptr.set* $\leftarrow$ *writeptr.set* $\cup \{i\}$
    **if** $|writeptr.set| = t + k$ **then**
        // Update metadata at *dir* with currently written value and with frozen values from previous write
        **invoke** $\langle$ *dir-Update* $\mid c,$ (*writeptr*, *frozenptrlist*, *frozenindex*, $*$) $\rangle$; **wait for** $\langle$ *dir-UpdateAck* $\rangle$
        // Obtain current reader indices
        **invoke** $\langle$ *dir-Scan* $\rangle$; **wait for** $\langle$ *dir-ScanResp* $\mid M$ $\rangle$
        *freets* $\leftarrow \{prevptr.ts\}$
        **forall** $p \in \mathcal{C} \setminus \{c\}$ **do**
            $(*, *, *, index) \leftarrow M[p]$
            **if** *index* > *frozenindex*$[p]$ **then**   // Client $p$ may be concurrently reading *prevptr* or *writeptr*
                *freets* $\leftarrow$ *freets* $\cup \{frozenptrlist[p].ts, reservedptrlist[p].ts\}$
                *frozenptrlist*$[p] \leftarrow$ *writeptr*; *frozenindex*$[p] \leftarrow$ *index*
                *reservedptrlist*$[p] \leftarrow$ *prevptr*
        *freets* $\leftarrow$ *freets* $\setminus \bigcup_{p \in \mathcal{C}} \{frozenptrlist[p].ts, reservedptrlist[p].ts\}$
        **forall** $j \in [1, n]$ **do**   // Clean up all fragments except for current, frozen, and reserved values
            **invoke** $\langle$ $d_j$-*Free* $\mid$ *freets* $\rangle$
        **invoke** $\langle$ *r-WriteAck* $\rangle$

**Algorithm 2.** Protocol AWE, atomic register instance $r$ for client $c$ (part 2).

**upon** ⟨ *r-Read* ⟩ **do**
    **forall** $i \in [1, n]$ **do** *readlist*$[i] \leftarrow \perp$
    *readindex* $\leftarrow$ *readindex* $+ 1$
    **invoke** ⟨ *dir-Update* | $c, (*, *, *, readindex)$ ⟩; **wait for** ⟨ *dir-UpdateAck* ⟩
    // Parse the content of *dir* and extract the highest timestamp, potentially frozen for $c$
    **invoke** ⟨ *dir-Scan* ⟩; **wait for** ⟨ *dir-ScanResp* | $M$ ⟩
    *readptr* $\leftarrow$ *highestread*$(M, c, readindex)$
    **if** *readptr.ts* $= (0, \perp)$ **then**
        **invoke** ⟨ *r-ReadResp* | $\perp$ ⟩
    **else** // Contact the data nodes to obtain the data fragments
        **forall** $i \in$ *readptr.set* **do**
            **invoke** ⟨ $d_i$-*Read* | *readptr.ts* ⟩

**upon** ⟨ $d_i$-*ReadResp* | *vts*, $v$ ⟩ such that *vts* = *readptr.ts* $\wedge$ *readlist*$[i] = \perp$ **do**
    **if** $v \neq \perp \wedge H(v) =$ *readptr.hash*$[i]$ **then**
        *readlist*$[i] \leftarrow v$
        **if** $\left|\{j | readlist[j] \neq \perp\}\right| = k$ **then**
            *readptr* $\leftarrow$ *Nullptr*
            *retval* $\leftarrow$ *Reconstruct*$_{k,n}$(*readlist*)
            **invoke** ⟨ *r-ReadResp* | *retval* ⟩

---

**Algorithm 3.** Protocol AWE, implementation of data node $d_i$.

**State**
    *data*$[ts] \in \Sigma^*$, initially $\perp$, for $ts \in$ *Timestamps*         // Stored data values indexed by timestamp

**upon** ⟨ $d_i$-*Write* | $ts, v$ ⟩ **do**
    *data*$[ts] \leftarrow v$
    **invoke** ⟨ $d_i$-*WriteAck* | $ts$ ⟩

**upon** ⟨ $d_i$-*Read* | $ts$ ⟩ **do**
    **invoke** ⟨ $d_i$-*ReadResp* | $ts$, *data*$[ts]$ ⟩

**upon** ⟨ $d_i$-*Free* | *freets* ⟩ **do**
    **forall** $ts \in$ *freets* **do**
        *data*$[ts] \leftarrow \perp$
    **invoke** ⟨ $d_i$-*FreeAck* | $ts$ ⟩

| Protocol | Communication cost | | Storage cost |
| --- | --- | --- | --- |
| | Write | Read | |
| ORCAS-A [14] | $(1+m)n\ell$ | $2n\ell$ | $n\ell$ |
| ORCAS-B [14] | $(1+m)n\ell/k$ | $2n\ell/k$ | $mn\ell/k$ |
| CASGC [10] | $n\ell/k$ * | $\infty$ | $mn\ell/k$ |
| CT [9] | $(n+m)n\ell/(k+t)$ | $\ell$ * | $n\ell/(k+t)$ * |
| HGR [19] | $n\ell/k$ * | $\infty$ | $mn\ell/k$ |
| M-PoWerStore [12] | $n\ell/k$ * | $n\ell/k$ | $\infty$ |
| DepSky [4] | $n\ell/k$ * | $n\ell/k$ | $\infty$ |
| AWE (Sec. 3) | $n\ell/k$ * | $(t+k)\ell/k$ | $2m^2 n\ell/k$ |

Table 2: Comparison of the communication and space complexities of erasure-coded distributed storage solutions. There are $m$ clients, $n$ data nodes, the erasure code parameter is $k = n - 2t$, and the data values are of size $\ell$ bits. An asterisk (*) denotes optimal properties.

With respect to storage complexity, protocol AWE freezes and reserves two timestamps and their fragments for each writer-reader pair, and additionally stores the fragments of the last written value for each writer. This means that the storage cost is at most $2m^2 n\ell/k$ bits in total. The improvement described in a remark of Section 3.3 reduces this to $2mn\ell/k$ in the best case.

Table 2 shows the communication and storage costs of protocol AWE and the related protocols. We use the wait-free semantics achieved by AWE and others as the base case; in CASGC [10] and HGR [19], a read operation concurrent with an unbounded number of writes may not terminate, hence we state their cost as $\infty$. In contrast to AWE, DepSky [4] is neither wait-free nor amnesic and M-PoWerStore [12] is not amnesic. It is easy to see that AWE performs better than most storage solutions in terms communication complexity.

## 5 Analysis

In this section we prove that protocol AWE, given by Algorithms 1–3, emulates an atomic read/write register and is wait-free.

Whenever the metadata directory *dir* contains an entry $ts = M[c].frozenptrlist[p].ts$ we say that timestamp *ts* is *frozen by c for p*. If *ts* is frozen by some *c* for any *p*, then *ts* is simply *frozen*. Furthermore, considering the state of writer *c*, a timestamp *ts* is said to be *retained by c for p* when either *frozenptrlist*[$p$].*ts* = *ts* (this includes that *ts* is frozen by *c* for *p*) or when *reservedptrlist*[$p$].*ts* = *ts* (which means that *ts* is reserved by *c* for *p*). A timestamp is *retained* by *c* when it is retained by *c* for some *p*. We call the timestamp $M[c].writeptr.ts$ the *written* timestamp of *c*.

**Lemma 1 (Frozen timestamps).** *At any time the timestamps that a client has frozen are no larger than its written timestamp. More precisely, for all $c, p \in \mathcal{C}$,*

$$M[c].writeptr.ts > M[c].frozenptrlist[p].ts.$$

*Moreover, during any dir-Update operation of c, the timestamp $M[c].writeptr.ts$ and all timestamps $M[c].frozenptrlist[p].ts$ may only increase.*

*Proof.* From Algorithm 1 it follows that for any client $c$, the timestamps stored in $M[c].writeptr.ts$ in successive *r-Write* operations of $c$ increase. From the same algorithm, it is clear that $M[c].frozenptrlist[p].ts$ is only updated through a *r-Write* operation of $c$, and is set to the written timestamp of the preceding *r-Write* operation of $c$, which is strictly smaller than the written timestamp stored in $M[c].writptr.ts$. The second inequality follows analogously. Thus, the values stored in $M[c].frozenptrlist[p].ts$ only increase. □

We define the *timestamp of a register operation o* as follows: (i) for an *r-Write* operation, the timestamp of $o$ is the value assigned to variable *writeptr.ts* during $o$; (ii) when $o$ is an *r-Read* operation, then its timestamp is the value assigned to variable *readptr.ts* by *highestread*. Note that the timestamp of an *r-Read* operation is $(0, \bot)$ if and only if $o$ returns $\bot$. Furthermore, we say that a value $v$ is *associated* to a timestamp $ts$ whenever the timestamp of the register operation that writes $v$ is $ts$.

According to *highestread*, the timestamp in the returned pointer may be frozen (taken from the *frozenptrlist* field of $M$) or written (taken from the *writeptr* field of $M$), but not both.

**Lemma 2 (Read frozen timestamp).** *If the timestamp ts of a r-Read operation $o_r$ by client c has been frozen for c by a client w, then w executes two r-Write operations concurrently to $o_r$, where the dir-Scan operation of the former r-Write operation $o_{w,1}$ and the dir-Update operation of the latter r-Write operation $o_{w,2}$ occur between dir-Update and dir-Scan operations of $o_r$. Moreover, the timestamp of the r-Read operation $o_r$ is ts, the one associated with the value written by $o_{w,1}$.*

*Proof.* From Algorithm 2 it follows that for *highestread* within $o_r$ to return a frozen timestamp, then, if $M$ is the metadata snapshot returned by the *dir-Scan* operation during $o_r$, it holds $M[w].frozenindex[c] = readindex$. This means that $w$ invoked *dir-Update* with the most recent value of *readindex* before the *dir-Scan* during $o_r$. To do that, $w$ must have detected the change of the *readindex* entry in $M[c]$ caused by $o_r$ through the *dir-Scan* operation invoked during $o_{w,1}$. From Algorithm 1, this can only be the operation through which $w$ wrote the value associated to *ts*. □

**Lemma 3 (Partial order).** *Let o and o′ be two distinct operations on register r with timestamps ts and ts′, respectively, such that o precedes o′. Then $ts \leq ts'$. Furthermore, if o′ is of type r-Write, then $ts < ts'$.*

*Proof.* We distinguish between two cases, depending on the type of $o$.

**Case 1:** If $o$ is of type *r-Write*, the claim follows directly from Lemma 1 and from the algorithm of the writer. In particular, if $o'$ is of type *r-Read*, then, if there is no concurrent *r-Write* operation of the same client $w$ as $o$, $ts$ is returned as written timestamp by the *readfrom* function when called for $w$ and reader of $o'$. In addition, if $o'$ runs concurrently with a *r-Write* of $w$, then one of the two hold: (i) *ts* (or a higher timestamp if many *r-Write* operations have intervened) is frozen for $o'$ and is returned by the *readfrom* operation invoked by *highestread* in $o'$ for $w$, (ii) *ts* (or a higher timestamp if many *r-Write* operations have intervened) has not yet been frozen by $w$, in which case a written timestamp greater or equal to *ts* (by Lemma 1) is returned by the *readfrom* operation invoked by *highestread* in $o'$ for $w$.

**Case 2:** If $o$ is of type *r-Read*, then let $ts^*$ be the maximum value of the timestamp field *ts* in a *writeptr* at the time when the *dir-Scan* operation invoked by $o$ returns. Note that *highestread* obtains *ts* as this maximum or as a frozen timestamp. Lemma 1 implies now that $ts \leq ts^*$.

We now show that $ts \leq ts'$ by distinguishing two cases. First, if $o'$ is of type *r-Write*, the writer calls *dir-Scan* after $o$ completes and determines the maximum value of the *ts* field in any *writeptr*. Then it increments that timestamp to obtain $ts'$. This ensures that $ts' > ts^* \geq ts$, as claimed.

Second, if $o'$ is of type *r-Read*, then $ts'$ may either have been a written timestamp or a frozen timestamp (at the time when the client obtained the response of its *dir-Scan*). If $ts'$ has been written, then $ts'$ is the maximum value of the *ts* field in any *writeptr*, which is at least as large as $ts^*$ by Lemma 1 and by the atomicity of *dir*.

Alternatively, if $ts'$ has been frozen by writer $w$, then Lemma 2 applies and shows that there exist two *r-Write* operations by $w$ that are concurrent to $o'$, of which the first writes the value associated to $ts'$. As such, if $ts_w$ is the timestamp returned by the *readfrom* function invoked by any *r-Read* operation $o$ that precedes $o'$ and for writer $w$, then $ts_w \leq ts'$. Since this can be extended to all writers, it holds that $ts \leq ts'$.

**Lemma 4 (Unique writes).** *If $o$ and $o'$ are two distinct operations of type r-Write with timestamps $ts$ and $ts'$, respectively, then $ts \neq ts'$.*

*Proof.* If $o$ and $o'$ are executed by different clients, then the two timestamps differ in their second component. If $o$ and $o'$ are executed by the same client, then the client executed them sequentially. By Lemma 3, it holds $ts \neq ts'$. □

**Lemma 5 (Integrity).** *Let $o_r$ be an operation of type r-Read with timestamp $ts_r$ that returns a value $v \neq \bot$. Then there is a unique operation $o_w$ of type r-Write that writes $v$ with timestamp $ts_w = ts_r$.*

*Proof.* Operation $o_r$ by client $c$ returns $v$ and is, thus, complete. This means that the client has processed $k$ events of type $d_i$-*ReadResp* from distinct nodes in a set $\mathcal{D}_k$; according to the protocol, the client has verified that the response from every $d_i \in \mathcal{D}_k$ contains a timestamp $vts_i$ and a fragment $v_i$ such that $vts_i = ts_r$ and $H(v_i) = readptr.hash[i]$.

According to the code, the value *readptr* is computed from a *writeptr* or a *frozenptr*[c] entry stored in the metadata directory *dir*. This pointer must have been computed during the write operation with timestamp $ts_w$ and was later stored in *dir* by the same client. Note that by Lemma 4, no other write has timestamp $ts_w$. From the algorithm of the writer, it follows that the entries in *readhash* were generated as hash values of the fragments, i.e., $readhash[i] = H(\bar{v}_i)$, where $\bar{v}_i$ for $i = 1, \ldots, n$ represent the erasure-coded fragments of some value $\bar{v}$.

Based on the check by the reader and the security property of the hash function, this means that $v_i = \bar{v}_i$ for all $i \in \mathcal{D}_k$. The completeness of the erasure code now implies that the reconstruction yields $\bar{v} = v$, the value associated to $ts_w$ and written by $o_w$. □

**Lemma 6 (Read concurrent with multiple writes).** *Consider an operation $o_r$ of type r-Read invoked by a reader $c$, with timestamp $ts_r$. At the time when $c$ determines $ts_r$ (by* highestread*), there are at least $k$ distinct correct data nodes that store a data fragment (different from $\bot$) under timestamp $ts_r$ and they do not free this fragment before $c$ completes $o_r$.*

*Proof.* Suppose that $ts_r = (sn, w)$ and the writer is client $w$. Consider a sequence $o_{w,1}, \ldots, o_{w,m}$ of r-Write operations executed by $w$ with respective timestamps $ts_{w,1}, \ldots, ts_{w,m}$, of which some are concurrent to $o_r$. Now consider the linearization of *dir* and let $o_{w,i}$ be the last one among these r-Write operations whose *dir-Update* (denoted by *dir-Update*$_{w,i}$) precedes the *dir-Update* operation of the reader during $o_r$ (denoted by *dir-Update*$_r$). Let *readindex* denote the reader's index at the time when $c$ invokes *dir-Update*$_r$.

W.l.o.g. suppose that *dir-Update*$_r$ follows at least one *dir-Update* operation that is triggered by an r-Write operation of $w$; furthermore, suppose that $w$ executes at least one more r-Write operation *dir-Update*$_{w,i+1}$ after *dir-Update*$_{w,i}$.

We claim that $ts_r = ts_{w,i} \lor ts_r = ts_{w,i+1}$. To show this, we distinguish four cases, considering the linearization of operations on *dir*. Let *dir-Scan*$_{w,i}$ denote the second invocation of *dir-Scan* during $o_{w,i}$, the one from which the writer takes *readindex*.

**Case 1:** Suppose that *dir-Update*$_r$ precedes *dir-Scan*$_{w,i}$; this means that $w$ detects the concurrent read $o_r$ during $o_{w,i}$, in the sense that $w$ updates its variable *frozenindex*[c] to *readindex*.

(Case 1.a) If the *dir-Scan* operation of the reader $c$ during $o_r$, denoted by *dir-Scan*$_r$, precedes *dir-Update*$_{w,i+1}$, then $c$ obtains $ts_r = ts_{w,i}$ as the highest timestamp stored in $M$ by the algorithm.

(Case 1.b) Otherwise, *dir-Scan*$_r$ follows *dir-Update*$_{w,i+1}$; then the reader $c$ obtains $M$ such that $M[w]$.*frozenindex*[c] is equal to *readindex* and $ts_r = M[w]$.*frozenptrlist*[c].$ts = ts_{w,i}$, according to *readfrom* in the protocol and because $M[w]$.*frozenindex*[c] is equal to *readindex*.

**Case 2:** Suppose that *dir-Update*$_r$ follows *dir-Scan*$_{w,i}$. This means that *dir-Update*$_r$ takes place between *dir-Scan*$_{w,i}$ and *dir-Update*$_{w,i+1}$ and $w$ detects the concurrent read $o_r$ only during $o_{w,i+1}$, after executing *dir-Scan*$_{w,i+1}$. The same two sub-cases may occur now.

(Case 2.a) If *dir-Scan*$_r$ precedes *dir-Update*$_{w,i+1}$, then $ts_r = ts_{w,i}$, analogous to Case 1.a.

(Case 2.b) Otherwise, *dir-Scan*$_r$ follows *dir-Update*$_{w,i+1}$ and the reader obtains $ts_r = ts_{w,i+1}$. To see this, suppose that (Case 2.b.i) *dir-Scan*$_r$ precedes the *dir-Update*$_{w,i+2}$ in the subsequent *r-Write* operation of $w$ or there is no such *r-Write*; then, the value *readindex* of $c$ remains greater than $M[w].frozenindex[c]$ and thus $c$ sets $ts_r = ts_{w,i+1}$. Alternatively (Case 2.b.ii), suppose that *dir-Scan*$_r$ follows *dir-Update*$_{w,i+2}$; then, according to the protocol, the writer has already set $M[w].frozenindex[c] = readindex$ during *dir-Update*$_{w,i+2}$ and $c$ sets $ts_r = ts_{w,i+1}$ analogous to Case 1.b.

Suppose the reader determines that $readptr.ts = ts_r$; then the correct nodes in $readptr.set$ store a fragment of the associated value because at least $t + k$ nodes in $readptr.set$ have sent a $d_i$-*WriteAck* for $ts_r$ to the writer. Accounting for the up to $t$ faulty nodes, at least $k$ correct nodes have once stored a fragment in $data[ts_r]$. It remains to argue why these nodes do not free this fragment before $c$ completes $o_r$.

In Case 1.a, the writer detects the concurrent read during $o_{w,i}$ and therefore excludes the data fragments associated to $ts_r$ from garbage collection for $c$, by setting $frozenptrlist[r].ts$ to $ts_r$ in its state. According to the logic of the protocol, $ts_r$ remains frozen and the corresponding fragments are retained at least until $c$ invokes a subsequent read operation.

In Case 2.a, almost the same happens during $o_{w,i+1}$, when the writer detects the concurrent read. The writer sets $reservedptrlist[r].ts$ to $ts_r$ in its state. Again according to the protocol, $ts_r$ remains reserved and the writer retains the corresponding fragments at least until $c$ invokes a subsequent read.

Intuitively, Cases 1.a and 2.a demonstrate why $w$ retains two values during a write (the one being written and the one written before): $w$ does not know which one of the two the reader is about to access.

In Case 2.b.i, if the writer detects the concurrent read during $o_{w,i+2}$, then it reserves and retains $ts_r$ and the claim follows analogously to Case 2.a.

In Cases 1.b and 2.b.ii, the reader accesses a frozen value. Again, according to the protocol, $ts_r$ remains frozen and is retained at least until $c$ invokes a subsequent read operation. The lemma follows. □

**Theorem 7 (Atomicity).** *Given a atomic snapshot object dir, protocol AWE emulates an atomic MRMW register r.*

*Proof.* We show that every execution $\sigma$ of the protocol is linearizable with respect to an MRMW register. By Lemma 5, the timestamp of a *r-Read* either has been written by some *r-Write* operation or *r-Read* returns $\bot$.

We first construct an execution $\tau$ from $\sigma$ by completing all operations of type *r-Write* for those values $v$ that have been returned by some *r-Read* operation. Then we obtain a sequential permutation $\pi$ from $\tau$ as follows: (1) order all operations according to their timestamps; (2) among the operations with the same timestamp, place the *r-Read* operations immediately after the unique *r-Write* with this timestamp; and (3) arrange all non-concurrent operations in the same order as in $\tau$. Note that concurrent *r-Read* operations with the same timestamp may appear in arbitrary order.

For proving that $\pi$ is a view of $\tau$ at a client $c$ w.r.t. a register, we must show that every *r-Read* operation returns the value written by the latest preceding *r-Write* that appears before in $\pi$ or $\bot$ if there is no such operation.

Let $o_r$ be an operation of type *r-Read* with timestamp $ts_r$ that returns a value $v$. If $v = \bot$, then by construction $o_r$ is ordered before any write operation in $\pi$. Otherwise, it holds $v \neq \bot$ and according to Lemma 5, there exists an *r-Write* operation $o_w$ that writes $v$ with the same timestamp. In this case, $o_w$ is placed in $\pi$ before $o_r$ by construction. No other *r-Write* operation appears between $o_w$ and $o_r$ because all other write operations have a different timestamp and therefore appear in $\pi$ either before $o_w$ or after $o_r$.

It remains to show that $\pi$ preserves the real-time order of $\sigma$. Consider two operations $o$ and $o'$ in $\tau$ with timestamps $ts_o$ and $ts'_o$, respectively, such that $o$ precedes $o'$. From Lemma 3, we have $ts' \geq ts$. If $ts' > ts$ then $o'$ appears after $o$ in $\pi$ by construction. Otherwise $ts' = ts$ and $o'$ must be an operation of type *r-Read*. If $o$ is of type *r-Write*, then $o'$ appears after $o$ since we placed each *r-Read* after the *r-Write* with the same timestamp. Otherwise, $o$ is a *r-Read* and the two *r-Read* operations appear in $\pi$ in the same order as in $\tau$ by construction. □

**Theorem 8 (Wait-freedom).** *Given an atomic snapshot object dir and assuming that $n \geq 2t + k$, protocol AWE is wait-free.*

*Proof.* As the atomic snapshot *dir* operates correctly, all its operations eventually complete independently of other processes. It remains to show that no *r-Write* and no *r-Read* operation blocks.

For a *r-Write* operation, the client needs to receive $t + k$ $d_i$-*WriteAck* events from distinct data nodes before completing. As there are $n$ nodes and up to $t$ may be faulty, the assumption $n \geq 2t + k$ implies this.

During a *r-Read* operation, the reader needs to obtain $k$ valid fragments, i.e., fragments that pass the verification of their hash value. According to Lemma 6, there are at least $k$ correct data nodes designated by *readptr.set* that store a fragment under timestamp $ts_r$ until the operation completes. As the reader contacts these nodes and waits for $k$ fragments, these fragments eventually arrive and can be reconstructed to the value written by the writer by the completeness of the erasure code. □

## 6 Conclusion

This paper has presented AWE, the first *erasure-coded* distributed implementation of a multi-writer multi-reader read/write storage object that is, at the same time: (1) asynchronous, (2) wait-free, (3) atomic, (4) amnesic, (i.e., with data nodes storing a bounded number of values) and (5) Byzantine fault-tolerant (BFT) using the optimal number of nodes. AWE is efficient since it does not use public-key cryptography and requires data nodes that support only reads and writes, further reducing the cost of deployment and ownership of a distributed storage solution. Notably, AWE stores metadata separately from $k$-out-of-$n$ erasure-coded fragments. This enables AWE to be the first BFT protocol that uses as few as $2t + k$ data nodes to tolerate $t$ Byzantine nodes, for any $k \geq 1$.

Future work should address how to optimize protocol AWE and to reduce the storage consumption for practical systems; this could be done at the cost of increasing its conceptual complexity and losing some of its ideal properties. For instance, when the metadata service is moved from a storage abstraction to a service with processing, it is conceivable that fewer values have to be retained at the nodes.

## Acknowledgment

We thank Alessandro Sorniotti, Nikola Knežević, and Radu Banabic for inspiring discussions during the early stages of this work. This work is supported in part by the EU CLOUDSPACES (FP7-317555) and SECCRIT (FP7-312758) projects.